\documentclass[10pt,letterpaper]{article}
\usepackage[top=0.85in,left=2.75in,footskip=0.75in]{geometry}

\usepackage{amsmath,amssymb}

\usepackage{changepage}

\usepackage[utf8x]{inputenc}

\usepackage{textcomp,marvosym}

\usepackage{cite}

\usepackage{nameref,hyperref}

\usepackage[right]{lineno}

\usepackage{microtype}
\DisableLigatures[f]{encoding = *, family = * }

\usepackage[table]{xcolor}

\usepackage{array}

\newcolumntype{+}{!{\vrule width 2pt}}

\newlength\savedwidth

\usepackage{setspace} 
\raggedright
\usepackage[aboveskip=1pt,labelfont=bf,labelsep=period,justification=raggedright,singlelinecheck=off]{caption}

\makeatletter
\renewcommand{\@biblabel}[1]{\quad#1.}
\makeatother

\usepackage{lastpage,fancyhdr,graphicx}
\usepackage{epstopdf}
\fancyheadoffset[L]{2.25in}
\fancyfootoffset[L]{2.25in}
\begin{document}
\vspace*{0.2in}

\begin{flushleft}
{\Large
\textbf\newline{Local media and geo-situated responses to Brexit: A quantitative analysis of Twitter, news and survey data} % Please use "sentence case" for title and headings (capitalize only the first word in a title (or heading), the first word in a subtitle (or subheading), and any proper nouns).
}
\newline
%ANONYMOUS FOR REVIEW
%Please check this is how you want your name displayed eventually and add orcid.
%Genevieve Gorrell 0000-0002-8324-606X
%Mehmet E. Bakir 0000-0002-3012-8713
%Luke Temple
%Diana Maynard 0000-0002-1773-7020
%Jackie Harrison
%J. Miguel Kanai
%Kalina Bontcheva 0000-0001-6152-9600


Genevieve Gorrell\textsuperscript{*}\textsuperscript{1},
Mehmet E. Bakir\textsuperscript{1},
Luke Temple\textsuperscript{2},
Diana Maynard\textsuperscript{1},
Jackie Harrison\textsuperscript{3},
J. Miguel Kanai\textsuperscript{2},
Kalina Bontcheva\textsuperscript{1}
\\
\bigskip
\textbf{1} Department of Computer Science, University of Sheffield, Sheffield, South Yorkshire, UK\\
\textbf{2} Department of Geography, University of Sheffield, Sheffield, South Yorkshire, UK\\
\textbf{3} Department of Journalism Studies, University of Sheffield, Sheffield, South Yorkshire, UK\\
\bigskip

% Use the asterisk to denote corresponding authorship and provide email address in note below.
* g.gorrell@sheffield.ac.uk

\end{flushleft}
\section*{Abstract}
Societal debates and political outcomes are subject to news and social media influences, which are in turn subject to commercial and other forces. Local press are in decline, creating a ``news gap''. Research shows a contrary relationship between UK regions' economic dependence on EU membership and their voting in the 2016 UK EU membership referendum, raising questions about local awareness. We draw on a corpus of Twitter data which has been annotated for user location and Brexit vote intent, allowing us to investigate how location, topics of concern and Brexit stance are related. We compare this with a large corpus of articles from local and national news outlets, as well as survey data,  finding evidence of a distinctly different focus in local reporting. National press focused more on terrorism and immigration than local press in most areas. Some Twitter users focused on immigration. Local press focused on trade, unemployment, local politics and agriculture. We find that remain voters shared interests more in keeping with local press on a per-region basis.

%\linenumbers

% Use "Eq" instead of "Equation" for equation citations.
\section*{Introduction}
\noindent 2016 saw the UK vote to leave the European Union, and Donald Trump elected as president of the USA, and with this came a rising concern about truthfulness in politics and the quality of information people have available to them (e.g. ~\cite{rose2017brexit}). A parallel issue relates to the way attention is drawn and shaped (e.g. ``attention economy''~\cite{harsin2015regimes}). As media and public figures compete for attention, we may fail to notice the cost in terms of reduced attention for the practical issues that have more impact on voters' lives. When the UK voted to leave the EU on June 23rd 2016, it was unsurprising that voters on the border between Northern Ireland and the Republic of Ireland voted to remain in the EU. The prospect of a customs border separating Irish people from jobs and family is a significant concern for them. However, in other regions of the UK the relationship between voting behaviour and local interests requires more analysis in order to understand local concerns and priorities. Cosmopolitan, affluent areas were more likely to vote to remain in the EU, whereas more rural and less affluent areas voted to leave, despite the importance of trade relationships with the EU~\cite{theconversation} and the fact that the majority of EU funding received by the UK goes to support agriculture and development, making them net financial beneficiaries~\cite{los2017mismatch}. The case of Sunderland, in the North East of England, perhaps illustrates that receiving funding does not equate to a pro-EU attitude~\cite{nytimes,rushton2017myth}.
%https://fullfact.org/europe/how-much-do-regions-uk-receive-eu-funding/
%https://fullfact.org/europe/uk-leaving-eu-wales/
%https://fullfact.org/europe/uk-leaving-eu-scotland/
%Further links can be found on the topic from these
%, and strong local press can be a source of influence on national coverage, improving representation~\cite{seaton2016brexit}

Close ties in offline communities increase resistance to propaganda and media-promoted ideas~\cite{seaton2016brexit}, and might have offered some ballast against the more extreme voices in the Brexit campaign. Yet local newspapers have faced numerous challenges over the last two decades due to concentration of media ownership, prioritization of shareholder interests above readers through a reduction in the numbers of journalists in local newsrooms, a decline in traditional advertising revenue which has resulted in cost cutting and redundancies, competition from companies such as Facebook and Google, which offer locally targeted advertising, the growth of social media as sources of local information, and changing news audience consumption patterns~\cite{ramsay2016monopolising}. The steady loss of traditional local print news sources through the closure or mergers of weeklies in local communities, daily newspapers in towns and cities and the demise of regional newspapers has led to ``news gaps''~\cite{currah2009navigating} (unreported areas of concern), raising questions about the means by which communities can stay informed about relevant local affairs, develop community connections or sustain local identities within a geographic proximity~\cite{hess2014geo}.  The range and scope of local news coverage contributes to its vulnerability, as local news can be both a vital arm of local democracy, but it can also be of poor quality, routine, trite and trivial, ``frequently terrible and yet also terribly important''~\cite{nielsen2015local}. Local coverage can range from exposing corruption in local public office to the mundane village fete, with relevance and importance coexisting alongside entertainment and a prurient fascination with local crime, human interest and lifestyle stories. Sources of news in the local setting are generally limited to a ``news beat'' which can lead to routine engagement with a narrow range of local sources such as local government and business and the police~\cite{harrison2006news}, missing the opportunity for individuals to express their views via traditional local media. Despite this decline, many important stories originate from the journalism that takes place at ground level in urban and rural areas and local news also tends to stay with stories once the national papers have left.

In this context, social media are both a problem and a possible solution. Whilst the success of digital local news and ultra-local start-ups have been patchy and audience take-up limited, the importance of local news has been recognized recently both by Facebook and the BBC. Facebook's Community News Project~\cite{facebook} supports local journalism in the UK by enabling the National Council for the Training of Journalists (NCTJ) to oversee the recruitment of eighty trainee journalists on a two year scheme. The £4.5 million investment aims at increasing local news reporting in areas of the UK that have lost their local newspaper. In 2017 the BBC allocated £8m a year to pay for 150 reporters to work for local news organizations rather than the BBC~\cite{bbc,nyt2} to cover council meetings and public services, both areas increasingly neglected by the local press, and to share these stories with the BBC. A 2018 OFCOM (UK broadcasting regulator) report of audiences’ use of news media indicates that 23\% of news users use printed local or regional newspapers, showing that despite its decline in recent years, local newspapers still remain an important source of independent journalism about local public affairs.
There is also evidence that online social networks have a strong local component, providing a new channel for localized discussion~\cite{bastos2018geographic}.
%If it is true that ``Brexit was won in the regions''~\cite{seaton2016brexit}, then the situated impact of social media cannot be ignored. 

In this work, we use multiple methods and multiple sources to scope these issues. A large corpus of local and national news articles is drawn on to investigate the way in which regional interests were covered in the Brexit debate period. We cross-compare this with a corpus of Twitter data in which Brexit is discussed, considering how interests are reflected and media are amplified. In this corpus, the location and Brexit stance of the tweeters have been ascertained, enabling a novel localized, vote-orientated analysis to be performed. We also use survey data to ground-truth our findings with a sample of the electorate, particularly with regards to media consumption. These three corpora taken together allow a new comprehensive picture to be formed, through quantitative investigation using topic modelling and named entity recognition. Specifically, our work addresses the following research questions:

\begin{itemize}
    \item What were the main foci in national and local newspaper coverage in the different regions of the UK in the run-up to the referendum, and how did they differ?
    \item How did newspaper coverage compare with what individuals from different regions were discussing on Twitter? What does this say about media representation of real interests?
    \item How do interest patterns and media consumption differ between those who favoured leaving the EU and those who favoured remaining in it in different parts of the country? How do they relate to local and national media?
\end{itemize}

\section*{Related research}

%%%Previous work on Twitter and Brexit and media and Twitter and Brexit%%%

%% https://journals.plos.org/plosone/article?id=10.1371/journal.pone.0206841
Social media are increasing in popularity as a way for people to access news~\cite{newman2017reuters}, and this has the potential for widespread impact on global politics. Social media have been widely observed to provide a
platform for fringe views~\cite{faris2017partisanship,silverman2015lies,barbera2015understanding,preoctiuc2017beyond}, and political asymmetries on social media are a further cause for unease. Research in the US political context~\cite{allcott2017social,silverman2015lies} raises concerns regarding polarized political debates in other countries. Ferrara~\cite{ferrara2017disinformation} presents findings on
the anti-Macron disinformation campaign in the run-up to the 2017 French presidential election, and a body of work~\cite{lansdall2016change,mangold2016should} has begun to explore Brexit.
%We contribute with a Twitter analysis of geolocated, vote-classified users and an emphasis on what we can learn about the changing relationship with news media. 

Bastos et al.~\cite{bastos2018geographic} present similar work to ours, in which a sample of Twitter users has been classified for Brexit vote intent and situated geographically. They explore network structure in mentioning and retweeting to explore the relationship between ``echo chambers'' and geography. They find that remainers often had links to people who were further away from them, whereas leavers tended to be more linked to people who were geographically closer to them. While our Twitter corpora are similar, our work differs from theirs in its focus on local and national media influences. We also include evaluation results for our stance and location classification.

Gorrell et al.~\cite{gorrell-et-al-2018} explore the media influences that dominated the Twitter Brexit debate. The most linked newspaper is the Guardian, which reflects the majority remain stance on Twitter. However, an aggressive minority of leavers tweeted more media links in total, most notably to mainstream media such as the Express as well as a plethora of alternative media producing materials attracting a strong leave audience. Upheld press complaints centred particularly on the Express and other leave media, supporting others' research regarding issues with truthfulness in the campaign~\cite{moore2017uk}. Compared with that work, the novel contribution here is the focus on situatedness, as well as the new insights enabled by topic modelling of news text. Locating the users geographically, as well as enabling investigation of how views relate to location, also makes it more possible to give some indicators about whether the impression created by Twitter materials is plausible as a reflection of the attitudes and media-related behaviours of the populace, or whether it is skewed by, for example, ``astroturfing'' (campaign accounts posing as voters in order to influence) or other distorting influences.

%% mention the ego-network stuff done in SoBigData a while back by Valerio Arnaboldi. (I don't have a reference but I have slides with it in - sale/internal/talks/clarin-social-media-may2017/ The conclusions were similar about remainers being more social than leavers, while leavers have more frequent interactions within a smaller circle. However, what we don't know is how this correlates with geography - we know that cosmopolitan dwellers tend to have wider networks than rural ones, so they correlate well with remainers, but are there actual causal relationships here?

Brexit media research is highly relevant here. Research undertaken by Loughborough University's Centre for Research in Communication and Culture showed the extent of press bias towards the referendum among main UK outlets, with the Financial Times and The Guardian in strong support to remain in the EU, and The Sun and Daily Mail supporting the leave campaign~\cite{deacon2016hard}. Additional reports from this centre indicate consistency in the issue agendas: both types of outlets covered referendum conduct, economy/business, and immigration as the three most prominent issues~\cite{lboro}. Previous analysis~\cite{nesta} has investigated attitudes towards particular Brexit campaign topics on Twitter. For example, issues related to employment were discussed much more frequently by remainers than leavers, while issues related to immigration and democracy were discussed much more frequently by leavers. We add to this through drawing connections with local and national news. Moore and Ramsay contrast tabloid and broadsheet coverage~\cite{moore2017uk}, and Matsuo and Benoit~\cite{lse} investigate differences in the dialogue between leave and remain camps, but little attention so far has focused on the difference between local and national coverage. This is an important novel contribution of our work.

More broadly, this research is positioned in the context of a global ``information malaise'', and concerns about the integrity of democracy. In the context of emerging geographical research on Brexit and underlying patterns of economically left-behind areas and the discontent of their socially disadvantaged populations \textit{vis-a-vis} metropolitan elitism~\cite{manley2017geography,dorling2016brexit}, more research is needed to ascertain the degree to which diverse publics were able to comprehend the regulatory complexities of scale and make full sense of its implications for their situated life chances. Los et al.~\cite{los2017mismatch} provide a comprehensive review, and the same authors provide an illustrative graph~\cite{cer} relating EU exports from a region with their EU membership attitudes. Such research acknowledges a regional aspect to attitudes, for example around Scottish identity, but according to Manley et al.~\cite{manley2017geography} geographical variation in Brexit attitudes can be elucidated as an expression of differences in population such as age and qualification. We reserve such an analysis for future work.

%% In our Understanding Society project, some findings which may or may not be interesting (these are unpublished, but presented in a meeting at US):
%- The Herald mentioned local issues a lot less than the Evening Standard did (as a proportion of all its mentions of local/global/national/regional issues), and both mentioned local issues a lot less than all the national papers. I'm not sure if that makes sense though, so perhaps those results are dodgy.
%- The Herald talked least positively about Europe, and the Evening Standard the most, out of all the newspapers.,

\section*{Corpora}

In this section, we describe the three main data sources used in the work. We begin with the news corpus, before discussing the Twitter corpus and the methods we used to ascertain location and Brexit stance. Thirdly we introduce the survey data we draw on. In the following section we present the topic modelling and entity detection approaches we use throughout the work to profile interests and concerns.

\subsection*{Newspaper corpus}

We collected articles from national and regional newspapers that are available in the Nexis (\url{https://www.nexis.com}) database. The selection of national newspapers was based on those with the highest circulation value that were available in Nexis, comprising a mix of tabloid and broadsheet papers. We collected articles including any of the keywords ``Brexit'', ``EU'', ``referendum'', or ``article 50'' in the body of the text that were published between February 20th and June 23rd 2016 (the date of the referendum announcement until the referendum itself). The list of newspapers and the number of matched articles are given in Tables 1, 2 and 3 of \nameref{appendix}. Unfortunately, the Express was not available for inclusion and could not be accessed in a methodologically consistent way. Our corpus is large enough, with over 35,000 articles spanning the political spectrum and including both tabloids and broadsheets, to form a fair representation of the UK media landscape despite this, and media material has not been analyzed on a publication-by-publication basis in this work.

These data have been used to form two region-sensitive divisions. Firstly, regional newspaper articles can be divided according to region of publication. Secondly, both regional and national articles can be divided according to the location that the article is primarily \textit{about}. We identified the location names mentioned in the articles by matching text to a gazetteer list of UK location names extracted from DBpedia (\url{https://wiki.dbpedia.org/}), a structured encyclopedia derived from Wikipedia and suitable for machine use. The DBpedia location entities are also assigned additional properties such as the coordinates of the locations. We used the coordinate information to identify the level 1 Nomenclature of Territorial Units for Statistics (NUTS, \url{https://ec.europa.eu/eurostat/web/nuts/background}) region of locations mentioned in the articles. In this way, a mention of ``Edinburgh'', for example, would be associated with the NUTS region ``UKM'', Scotland. The most frequently occurring NUTS region in each article was assigned to it. Not all articles mentioned a UK location, so some articles were unassigned. Table 4 of \nameref{appendix} gives article counts per region.

\subsection*{Twitter corpus}

Around 17.5 million tweets were collected using Twitter's Streaming API from 3rd April until 23 June 2016. The highest daily volume was 2 million tweets on June 23rd (only 3,300 were lost due to Twitter rate limiting - information that is available via the API), with just over 1.5 million during poll opening times. June 22nd was second highest, with 1.3 million tweets. The 17.5 million tweets were authored by just over 2 million distinct Twitter users (2,016,896). The tweets were collected based on the following keywords and hashtags: \textit{votein, yestoeu, leaveeu, beleave, EU referendum,  voteremain, bremain, no2eu, betteroffout, strongerin, euref,  betteroffin, eureferendum, yes2eu, voteleave, voteout, notoeu,   eureform, ukineu, britainout, brexit, leadnotleave}. These were chosen for being the main hashtags, and are broadly balanced across remain and leave hashtags.

Almost half a million of these users were able to be classified by Brexit vote intent, on the basis of tweets authored by them and identified as being in favour of leaving or remaining in the EU. Partisan hashtags such as ``\#voteleave'' at the end of a tweet quite reliably summarize the tweeter's position with regards to the referendum. The methodology used is described in more detail in Gorrell et al.~\cite{gorrell-et-al-2018}
The end result is a list of 208,113 leave voters and 270,246 remain voters, classified with an accuracy of 0.966. Accuracy of the approach was able to be ascertained using a test corpus derived from formulaic, explicit vote declarations generated in response to a campaign of the time.
%Above accuracy easily calculated from the Soc Info paper

Users were allocated to a NUTS1 region on the basis of text in the Twitter location field. This is a free text field that users may fill in in any way they choose, or choose not to fill in. As a result of this, users may ignore  the field, repurpose it or use it humorously, so only a limited number of locations could be identified reliably. In total, 162,548 user locations were obtained using the same approach as for the newspaper articles, in which text is matched to a gazetteer of UK location names from DBpedia, and the coordinates given are used to assign a NUTS1 region. In addition, a very small number of Twitter users (0.18\% of our sample) had consented to location coordinates being added to their tweets. These are too small in number to make an impact on the size of the dataset, but made it possible to tune and evaluate the work to some degree. The evaluation on a test set of 1016 users with location coordinates achieved a precision of 0.82, a recall of 0.67 and an F1 of 0.74. The actual accuracy is probably a little higher than this, since the coordinates may not be especially more reliable than the location field resolution, as users may have moved around.

\subsection*{Survey data}

In order to add context to this study, we profile both the general Twitter user and those who claim to 'share political information' on the platform by making use of wave 12 data from The British Election Study 2014-2018~\cite{bes}. The study is managed by a consortium of the University of Manchester, the University of Oxford and the University of Nottingham, and wave 12 was conducted by the market research firm YouGov between the 5th of May 2017 and the 7th of June 2017. 34,464 of the address-sampled respondents participated. The questionnaire covers a broad variety of demographic information, as well as politically relevant behaviours and attitudes including social media use.

\section*{Methods}

Two main natural language processing methods have been used to explore the subject matter in the news and tweet corpora in a quantitative way; a topic view and an entity view.

\subsection*{Topic modelling}

Latent Dirichlet Allocation (LDA)~\cite{blei2003latent} was used to discover topics discussed in the Twitter and newspaper corpora. In LDA, word frequencies in texts are considered to arise from a weighted mixture of latent topics. Topics are discovered automatically, having manually specified the desired number. Deciding on the number of topics is often done by plotting the coverage provided by the topics against a range of topic numbers; an ``elbow'' in the graph shows where increasing the number of topics starts to give a diminishing return in covering the data. However, since the nature of our work is to explore meaningful subject areas from a narrative point of view, we were obliged to increase the number of topics. The reader should bear in mind as we proceed that varying the topic number can have an unpredictable effect on outcomes, since for example ambiguous tokens (i.e. words or numbers) may move in and out of topics. It is for this reason that we use two methodologies (topic modelling and named entity extraction) in the work.

The final set of 75 topics was derived from the complete news corpus, as this provided a large enough quantity of material to extract detailed, high quality topics. The topics discovered covered a broad range of relevant subjects, including energy (0.8\% of tokens) and the ``Queen Backs Brexit'' story (0.7\% of tokens). Higher token coverage does not necessarily indicate a better topic; the two biggest topics (7.4\% and 6.9\% of tokens respectively) capture mainly the background language distribution (five most salient words in these with $\lambda=0.5$~\cite{sievert2014ldavis} are ``get'', ``like'', ``think'', ``thing'', ``know'' and ``even'', ``may'', ``seem'', ``yet'', ``political'', respectively, for illustration), and it is evident from the examples given that small topics can be very precise. This same set of topics was then applied to Twitter material as well as news subsets as necessary, in order to enable comparisons to be made between the differing datasets.

Of these topics, a number were pre-selected, guided by theory, in order to reduce the possibility of type one errors that would arise in calculating statistical significance of relationships across such a large number of topics. The theory was that local coverage would emphasise practical matters of relevance to local people's daily lives, whereas national coverage would emphasise matters such as national identity. We thus identified a range of topics that were likely to provide a good opportunity to investigate these points, as shown in Table~\ref{tab:topics}.

\begin{table}
\resizebox{.95\textwidth}{!}{%
\begin{tabular}{l|l|l|l}
\textbf{Topic} & \textbf{Five most salient tokens} & \textbf{\% of tokens} & \textbf{Theory}\\
\hline
Trade & trade single market free britain & 3.1 & Most Localities\\
Employment & work development need fund skill & 2.6 & Most Localities\\
Local politics & city council local town park & 1.0 & Most Localities\\
\hline
Steel & steel tata javid talbot port & 0.8 & Specific Localities\\
Agriculture & farmer farm rural managing broadband & 0.6 & Specific Localities\\
Car pollution & car pollution vehicle air diesel & 0.5 & Specific Localities\\
Fishing & fish water vessel fisherman fishery & 0.3 & Specific Localities\\
\hline
Immigration & immigration migration migrant wage immigrant & 1.9 & National Interest\\
Terrorism & attack brussels airport belgian metro & 1.8 & National Interest\\
Scotland & scotland scottish snp independence sturgeon & 1.1 & National Interest\\
Northern Ireland/Wales & ireland northern irish wales belfast & 0.5 & National Interest\\
\hline
\end{tabular}%
}
\caption{Topics selected for study, along with five most salient words (note that the topic was titled and selected based on the full set of words in the topic, not just the five shown). The percentages of tokens in the corpus covered by the topics are given, alongside our theoretical expectations regarding national relevance.}
\label{tab:topics}
\end{table}

\subsection*{Entities}

In the entity view, \textsc{TagMe} (\url{https://tagme.d4science.org/tagme/})~\cite{DBLP:conf/cikm/FerraginaS10} has been used to find mentions of entities in the corpora. An ``entity'' might be a person, place or organization, or it might be a concept; \textsc{TagMe} matches anything with a Wikipedia page. Any annotation has an associated value $\rho \in [0,1]$  which estimates the ``goodness'' of the annotation with respect to the other entities of the input text. We used this value to filter poor annotations, namely those with $\rho < 0.10$ (We used this value  as suggested in the documentation).

We then derived the entities associated with a newspaper as the aggregation of entities found in its articles. Using the correspondence between articles and NUTS1 regions described above, we also derived regional entities as the aggregation of annotations found in articles with same associated region.

In order to effectively  extract the ``key'' topics on both aggregates, we exploited the following scoring technique to rank entities:

\begin{itemize}  
%\item $score(e, A) = articles\_count(e, A) * mean\_\rho(e, A)$
\item $score(e, A) = articles\_count(e, A) * max\_\rho(e, A)$
%\item $score(e, A) = articles\_count(e, A)$
\end{itemize}

where  $articles\_count(e,A)$ is the number of articles in the aggregate $A$  containing entity e and $max\_\rho(e, A)$  is the max value of $\rho$ among the occurrences of entity $e$ in $A$. The final rank is obtained by ordering entities according to their scores.

%where  $articles\_count(e,A)$ is the number of articles in the aggregate $A$  containing entity e, $mean\_\rho(e, A)$ (respectively $max\_\rho(e, A)$)  is the mean (respectively max) value of $\rho$ among the occurrences of entity $e$ in $A$. The final rank is obtained by ordering entities according to their scores. The first two approaches gave similar results, while the last one resulted in a worse performance, perhaps because it does not consider the $\rho$ score associated with entities.
%, so any common wrong annotation ends up having a high score and thus a high final rank.
%A clear example of this is the entity \textit{Will\_and\_testament} associated to the spot will. This clearly wrong annotation was present in the vast majority of articles with an associated $\rho$  always close to $0.1$, i.e. the threshold we used to filter out entities. We were able to increase the goodness of results by using the first two approaches.

The entity view benefits from \textsc{TagMe}'s ability to disambiguate mentions of entities. For example, Theresa May might be referred to as ``Mrs. May'', or ``the UK prime minister''. Therefore, counts of mentions can be grouped across different ways of expressing a concept. It also makes it easier to focus on important entities, as only entities salient enough to have a Wikipedia page are extracted. It is like asking closed questions; how do people talk about, for example, Brussels? Trade tariffs? Who talks most about these subjects? The topic view is more like asking open questions; we take our lead from what the texts themselves focus on. This is important as closed questions can miss unexpected trends or more subtle effects. For example, subtleties such as trends toward nationalism in the discourse may be lost in the entity view but might appear as a topic using LDA. It was possible to select baskets of entities that matched the pre-selected LDA topics well, enabling various forms of parallel analysis.

\section*{Ethics}

Ethics approval was obtained for the Twitter data collection from the University of Sheffield (application number 011934), and Twitter's terms and conditions for the collection of such data are complied with. Only publicly shared data are used; however, due to the sensitive nature of political opinion, the data are held confidentially and for a limited time duration, as required by the ethics committee. Anonymous aggregate data are available at \url{https://gate-socmedia.group.shef.ac.uk/wp-content/uploads/2020/04/brexit-geomedia-shared-data.zip}

\section*{Results}

We discuss each research question in turn. First we review newspaper coverage, considering both overall emphasis and regional foci, as well as contrasting local and national coverage, and impressions related to particular areas. We then explore reception via the Twitter corpus, considering regional differences. Finally we relate findings on a per-region basis to vote declarations on Twitter as well as referendum outcomes.

\subsection*{RQ1: National and local newspaper coverage}

%%Begin the section setting up topics for focus throughout the rest of the paper, and talking about how local and national coverage differed on a national level.

As previously mentioned, topics were derived on the entire newspaper corpus, and used throughout the work.
%The relative magnitude of the topics, quantified as the percentage of tokens (semantic units of text, usually words) explained by that topic, allows us to get a sense of what were the main foci of interest in the newspaper corpus as a whole.
The two largest topics covered background language distributions, enabling other topics to uncover semantically coherent areas in contrast. The third most dominant topic was Brexit itself, which is unsurprising given that the corpus was deliberately comprised of articles mentioning Brexit. The fourth topic is background language relating to the business of government and law, and after that we see topics that give a sense of what the media considered to be relevant issues to Brexit. The economy and the government feature highly, as do aspects of the Brexit campaign, supporting Loughborough's findings~\cite{deacon2016hard} about press focus. 

The top 14 of these are given in Table~\ref{tab:toptops}. Further background language or unclear topics have been excluded as uninformative.

\begin{table}
\begin{center}
\resizebox{.75\columnwidth}{!}{%
\begin{tabular}{l|l|l|l}
\textbf{Topic} & \textbf{\% Toks} & \textbf{Topic} & \textbf{\% Toks}\\
\hline
Trade (UK internat.) & 3.1 & Refugees (Calais) & 1.9\\
The economy & 2.9 & Football & 1.9\\
David Cameron & 2.8 & Business & 1.8\\
Brexit campaign & 2.8 & Terrorism (Brussels) & 1.8\\
Employment & 2.6 & UK treasury & 1.8\\
UK politics & 2.2 & Russia/Ukraine & 1.7\\
Immigration & 1.9 & EU politics & 1.7\\
\hline
\end{tabular}%
}
\caption{Most significant topics in newspaper corpus}
\label{tab:toptops}
\end{center}
\end{table}

\begin{figure}[h!]
  \centering
    \includegraphics[width=.9\textwidth]{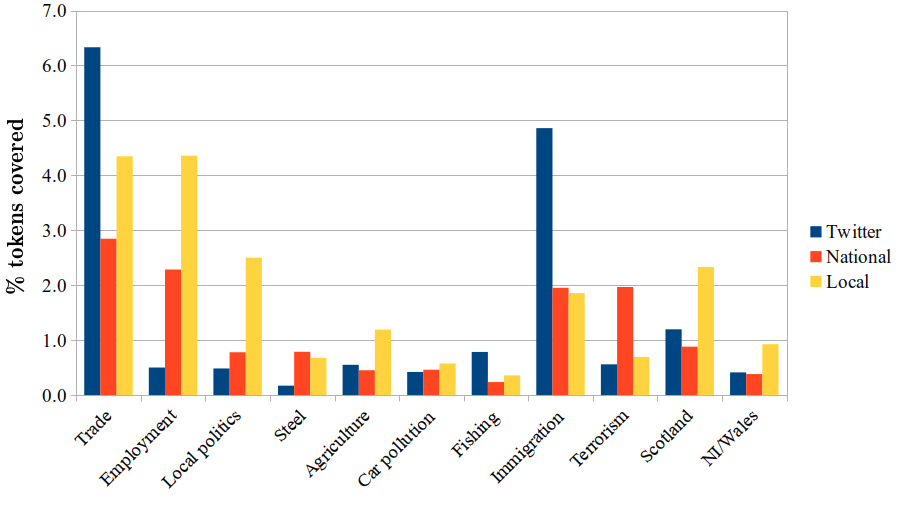}
    \caption{Topics (from topic modelling) - percentage of tokens covered by selected topics in national and regional press vs. Twitter, showing differing emphases.}
\label{fig:topics-histo}
\end{figure}

In terms of topics, national papers talk about the economy, David Cameron, trade, employment, refugees, terrorism, and immigration. Regional papers discuss employment, trade, football, the economy, UK politics, local politics, and Scotland. Notable differences from national coverage include a greater emphasis on employment, football, local politics and Scotland, and a reduced emphasis on terrorism. Fig~\ref{fig:topics-histo} shows the extent of representation of our pre-selected topics in national and local news articles, alongside Twitter findings which will be discussed later in the work.

\subsubsection*{Regional findings}

\begin{figure}[h!]
  \centering
    \includegraphics[width=.9\textwidth]{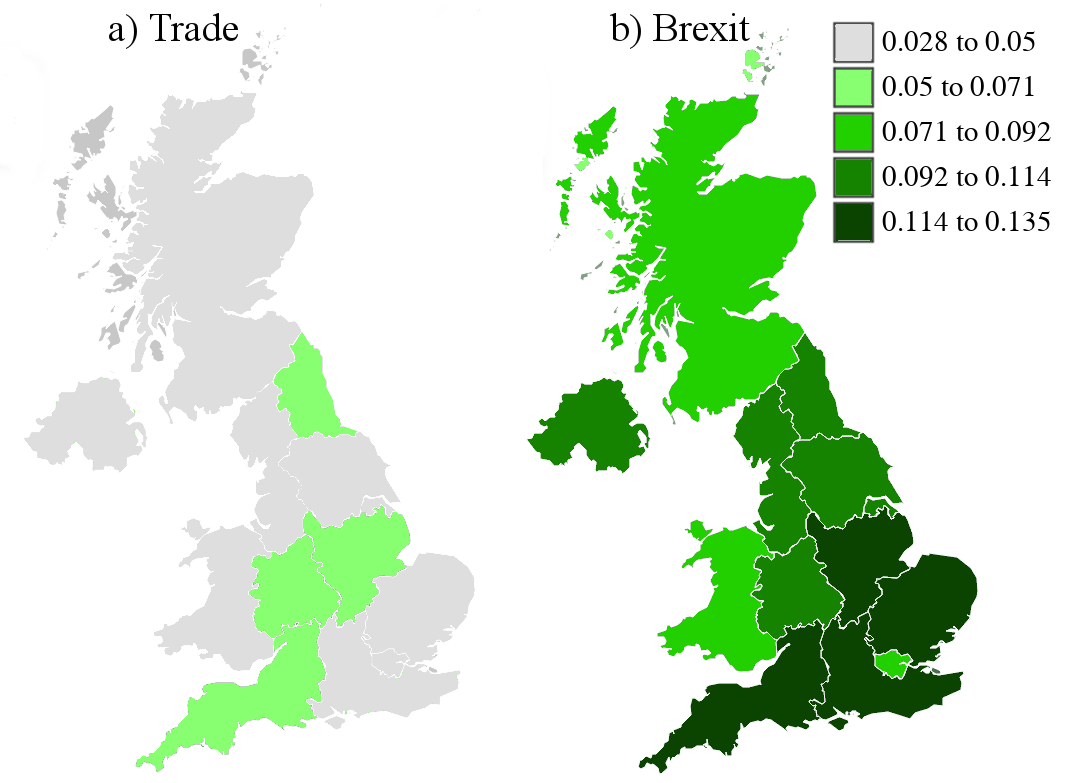}
    \caption{Topics mentioned by local papers published in regions. Proportion of topic tokens on average per article are shown, scaled to the maximum value across the topics in the figure. Created using JSON map data shared with the authors under a CC BY license, with permission from Mark Greenwood.}
\label{fig:tops-reg}
\end{figure}

In order to explore how local interests and local foci were reflected in the national narrative, we looked at topic representation in the different regions. Regional variation in interests was found. For example, agriculture was mentioned most in local papers in East Anglia. Steel was mentioned most by local papers in Wales, as well as the North East, and most by national papers in conjunction with Wales. Choropleths for all selected topics are available on the project website (\url{http://services.gate.ac.uk/politics/ba-brexit/}) and examples of trade and Brexit mentioning in local papers are given in Fig~\ref{fig:tops-reg}. Darker green shade indicates more topic representation in that region. Brexit mentioning pattern in local papers suggests that a focus on Brexit may have disposed toward voting leave, or reflected local leave tendencies, a hypothesis that is tested below under research question 3.

We now compare topics emerging from regional publications with the overall picture from national papers. It was hypothesized that local interests would be more practical, and that employment, trade and local government would be more widely mentioned in local publications than nationally. Steel, agriculture, car pollution and fishing were expected to be mentioned more than in national publications only in certain regions. Conversely, it was predicted that immigration, terrorism, Scotland and Northern Ireland/Wales would be mentioned more in national publications (other than, in the case of Scotland and Northern Ireland/Wales, in the regions themselves).

\begin{figure}[h!]
  \centering
    \includegraphics[width=.9\textwidth]{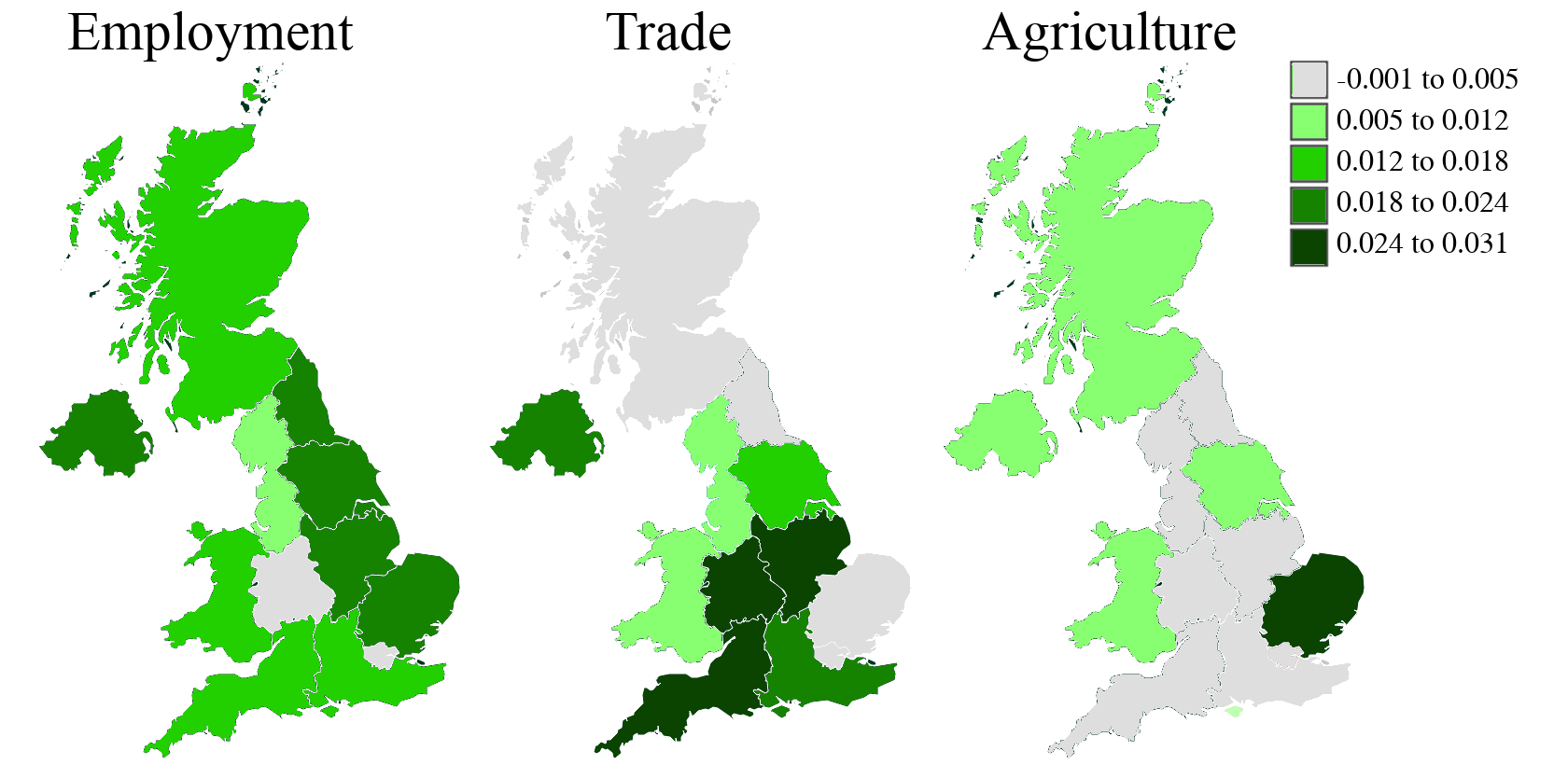}
    \caption{Topic emphasis in local vs. national newspapers. Green indicates that the topic is mentioned more in local papers. Created using JSON map data shared with the authors under a CC BY license, with permission from Mark Greenwood.}
\label{fig:tops-vs-nat}
\end{figure}

It was confirmed that employment was mentioned significantly more in most regions than nationally, as was trade. Agriculture was found to be of more widespread interest than had been predicted, but was indeed more often mentioned in local publications, as shown in Fig~\ref{fig:tops-vs-nat}. Local politics was also mentioned more in regional publications. In the figure, national coverage has been subtracted from local coverage (proportion of tokens covered by the topic, averaged over articles) to show regional differences in the contrast with national coverage, grey indicating more similarity. All differences are significant at p\textless0.001 in a Mann Whitney test.

\begin{figure}[h!]
  \centering
    \includegraphics[width=.9\textwidth]{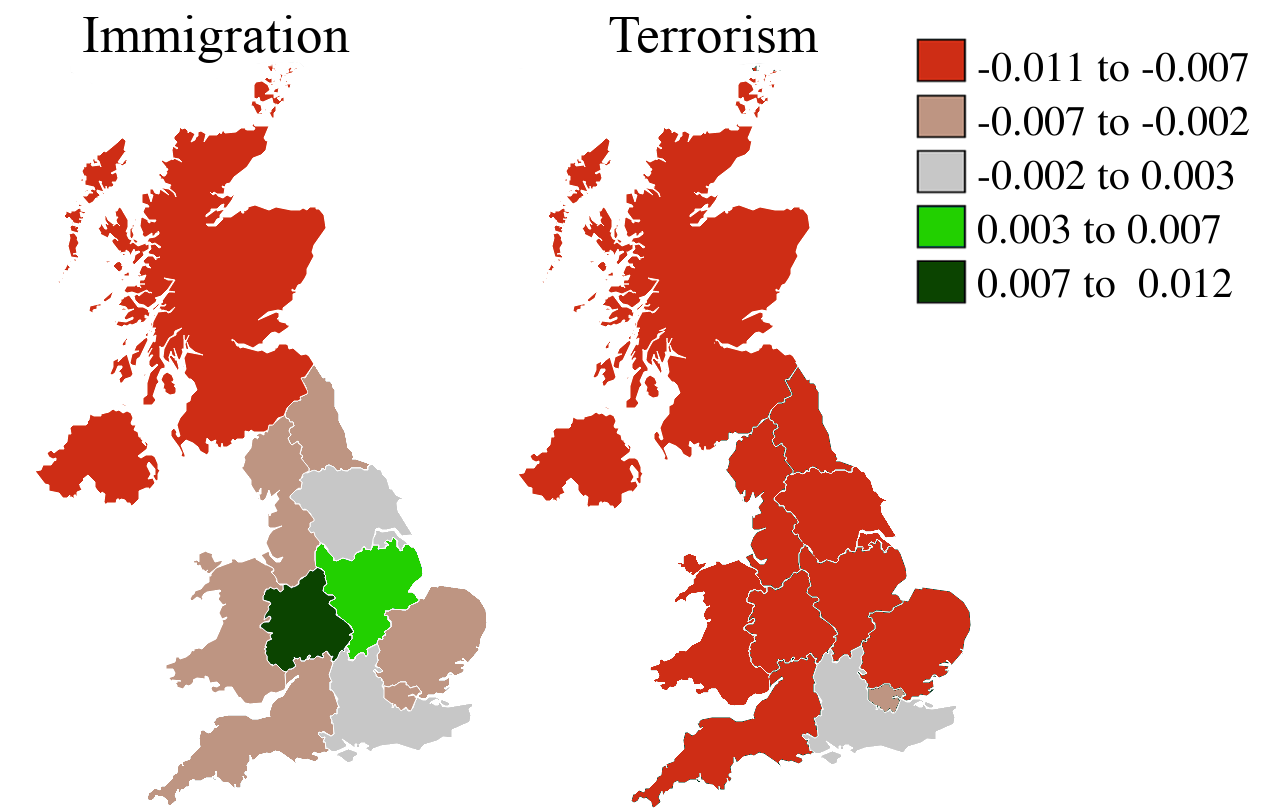}
    \caption{Topic emphasis in local vs. national newspapers; further topics. Red indicates that the topic is mentioned more in national papers, and green, less. Created using JSON map data shared with the authors under a CC BY license, with permission from Mark Greenwood.}
\label{fig:tops-vs-nat-more}
\end{figure}

Fig~\ref{fig:tops-vs-nat-more} shows that terrorism was widely mentioned significantly less in local papers than national. Immigration is somewhat more of a national topic than a local concern, though the picture is mixed. The figure shows topic representation in local papers minus in national papers, and all findings are significant at p\textless0.001 in a Mann Whitney test. Steel also presents a varied picture, with more coverage in Wales, the North East and in Yorkshire and the Humber, and less coverage in other regions. This is unsurprising, given steel industry location around the country. Car pollution was mentioned more in all regions except Yorkshire and the Humber, the East of England, Scotland and Wales. Patterns of interest might be seen as reflecting local concerns. Fishing was mentioned more in all regions but the North West, London and Wales. Topics mentioned less in local coverage are the regional topics of Scotland and Northern Ireland/Wales, where interest was limited to the regions themselves. Again, all findings are significant at p\textless0.001.

Data are too voluminous to give in full but Table~\ref{tab:tops-all-regions} gives regional/national differences for all regional papers aggregated, though note that this glosses over important regional differences as described above, and not all regions are equally represented. High significances arise from the large sample sizes. All significance testing was performed using two-tailed Mann Whitney tests on a per-article basis. In summary, findings broadly support our hypothesis that local and national news coverage differs, that local coverage emphasizes different topics, and that regions have their own distinct interests.

\begin{table}
\resizebox{.65\textwidth}{!}{%
\begin{tabular}{l|l|l|l|l}
\textbf{Topic} & \textbf{Nat. av.} & \textbf{Reg. av.} & 
\textbf{p-value} & \textbf{Sig.}\\
\hline
Employment & 0.021 & 0.034 & 0.000 & ***\\
Agriculture & 0.005 & 0.011 & 0.000 & ***\\
Immigration & 0.06 & 0.025 & 0.000 & ***\\
Terrorism & 0.015 & 0.006 & 0.000 & ***\\
Scotland & 0.01 & 0.022 & 0.000 & ***\\
EU politics & 0.177 & 0.151 & 0.000 & ***\\
Trade & 0.028 & 0.042 & 0.000 & ***\\
Local government & 0.007 & 0.02 & 0.000 & ***\\
NI/Wales & 0.005 & 0.009 & 0.000 & ***\\
Steel & 0.007 & 0.007 & 0.000 & ***\\
Fishing & 0.003 & 0.004 & 0.000 & ***\\
Car pollution & 0.006 & 0.006 & 0.000 & ***\\
\hline
\end{tabular}%
}
\caption{Mann Whitney tests comparing average topic mentions per article in national media vs local media. First two columns give the average number of mentions per article. The p-value and a visual significance indicator are then given.}
\label{tab:tops-all-regions}
\end{table}

As mentioned in the corpora section, national press articles can be grouped according to the region the article is \textit{about}, as ascertained through location mentions in the article. Mentioning regions and the issues that are important to them is a critical way for national press to show local sensitivity, perhaps even more than by reflecting their interests in the national discourse without mentioning the region. Topic representations in national press articles, divided into regions according to location mentions, were correlated with topics in local press coverage per-region to determine the extent to which emphasis agrees. The strongest correlations were for the topics of Scotland (0.99, p\textless0.001) and Northern Ireland/Wales (0.96, p\textless0.001), as might be expected given that the topics would themselves be annotated as location mentions. After that, strong correlations were found for steel (0.83, p\textless0.001) and immigration (0.71, p\textless0.001). Local politics correlates significantly (0.79, p\textless0.01), perhaps because regions with their own topic (Scotland, Northern Ireland and Wales) seem to have a much reduced focus on local politics, so this could be seen as the ``negative'' of the strong correlations we see for those regional topics.

Aside from the above, we do not see strong correlations for the other topics, which means that national reporting does not show a high degree of reflection of regional differentiation by topic. Agriculture, car pollution and fishing show weak, non-significant correlations, while terrorism, trade and employment show weak \textit{negative} correlations.

\subsubsection*{Entities}

\begin{figure}[h!]
  \centering
    \includegraphics[width=.9\textwidth]{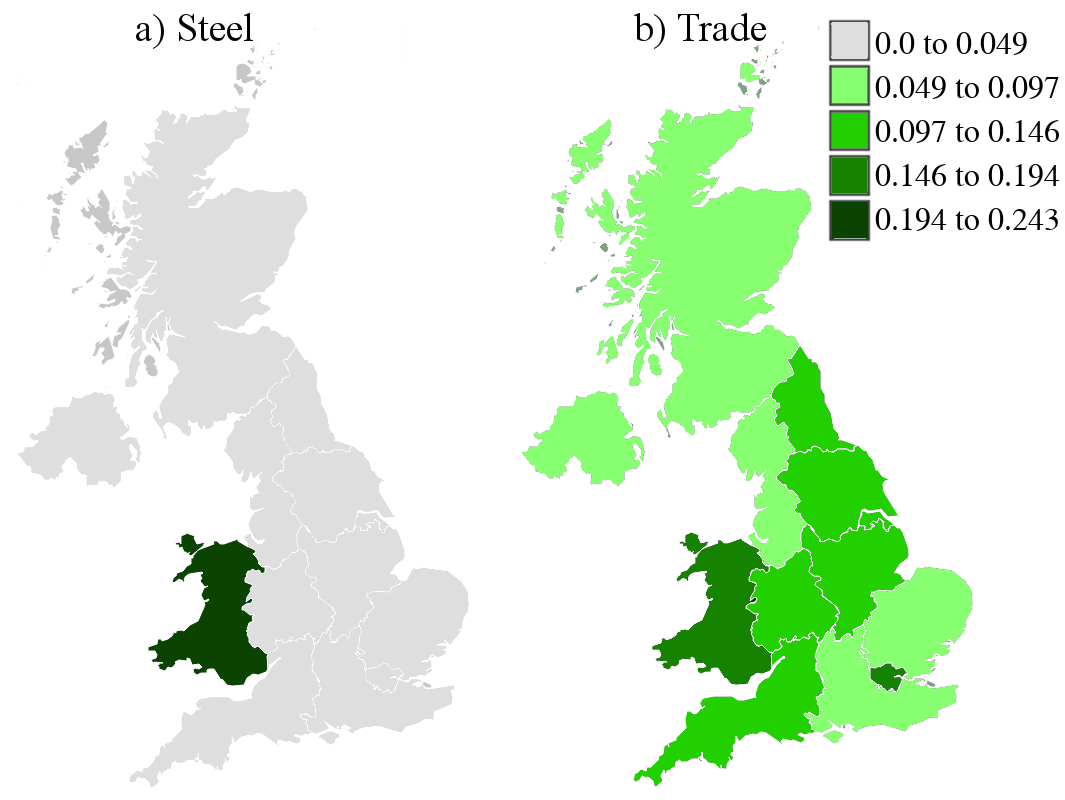}
    \caption{Entities mentioned by national papers (average mentions per article) in connection with regions. Values are scaled to the maximum value of the figure. Created using JSON map data shared with the authors under a CC BY license, with permission from Mark Greenwood.}
\label{fig:ents-nat}
\end{figure}

A parallel analysis was performed using entity mentions, on a per-newspaper basis. Matched entities were used for the pre-selected topics, as mentioned above. Entities were associated with regions in a similar manner to that described above for topics; i.e. by associating articles with regions that are mentioned in them (or places in those regions). However in order to maximize the data, since entity mentions are more sparse, the entire corpus was used to associate entities with locations according to their being mentioned in the article, including regional newspaper articles. Representation is calculated as the average number of entity mentions for the subject in the article. Observe that the subject of trade is widely associated with a variety of regions, confirming the observation for topics, as shown in Fig~\ref{fig:ents-nat}. Darker shades indicate a higher incidence of entity mentioning; findings are scaled against the maximum score for the figure.

\begin{figure}[h!]
  \centering
    \includegraphics[width=.9\textwidth]{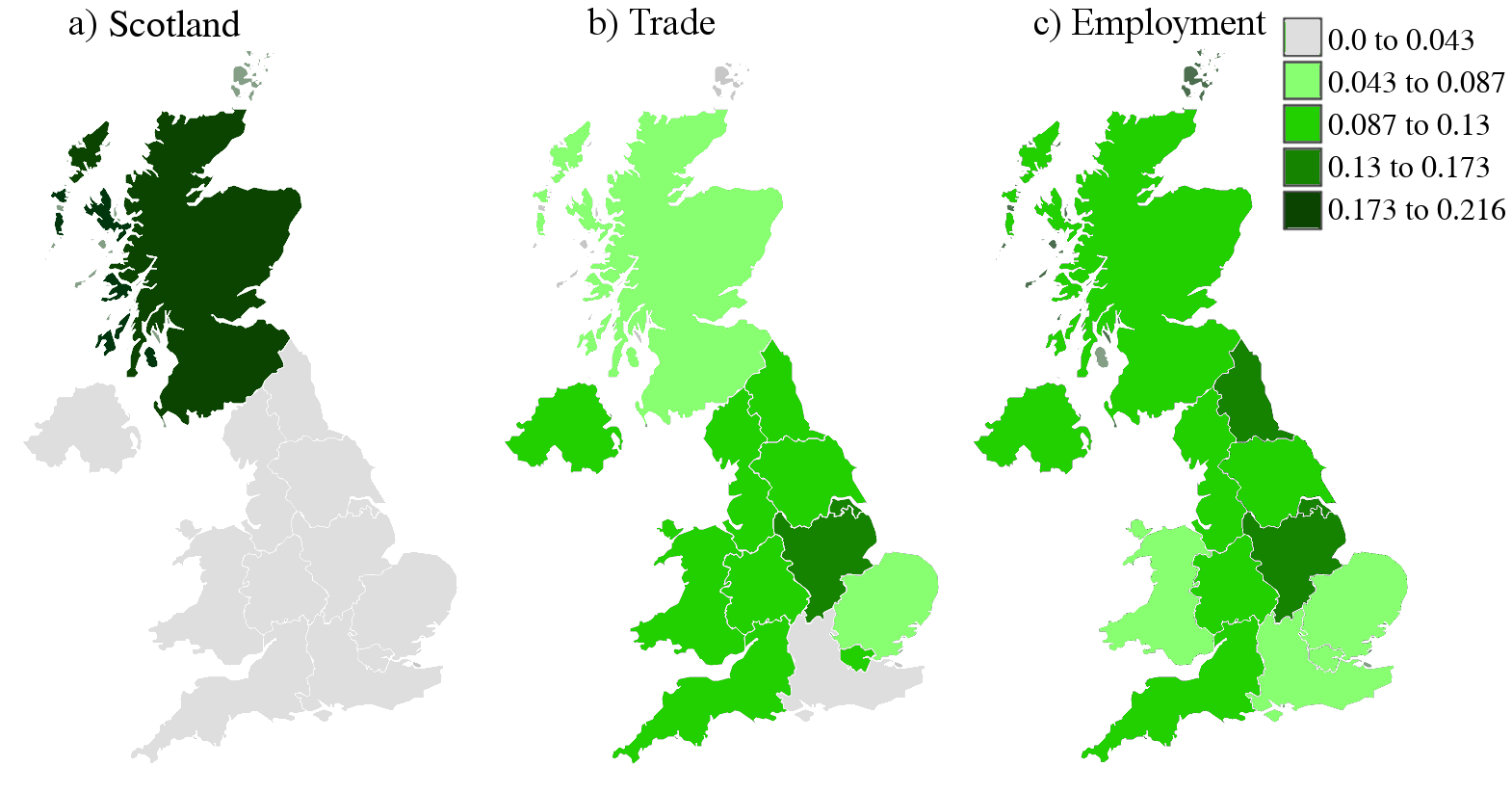}
    \caption{Entities mentioned by local papers in regions (average mentions per article) scaled to the maximum value for the figure. Created using JSON map data shared with the authors under a CC BY license, with permission from Mark Greenwood.}
\label{fig:reg-ents}
\end{figure}

Fig~\ref{fig:reg-ents} shows example entities discussed in regional papers. Trade is mentioned mainly in areas that voted leave, as above, but also in London and Northern Ireland. Employment was widely mentioned, suggesting substantiation for our hypothesis that employment was more important in regional papers than national ones. This is further explored in the subsection on research question 3 below. Again, further choropleths can be found on the project website.

\begin{figure}[h!]
  \centering
    \includegraphics[width=.9\textwidth]{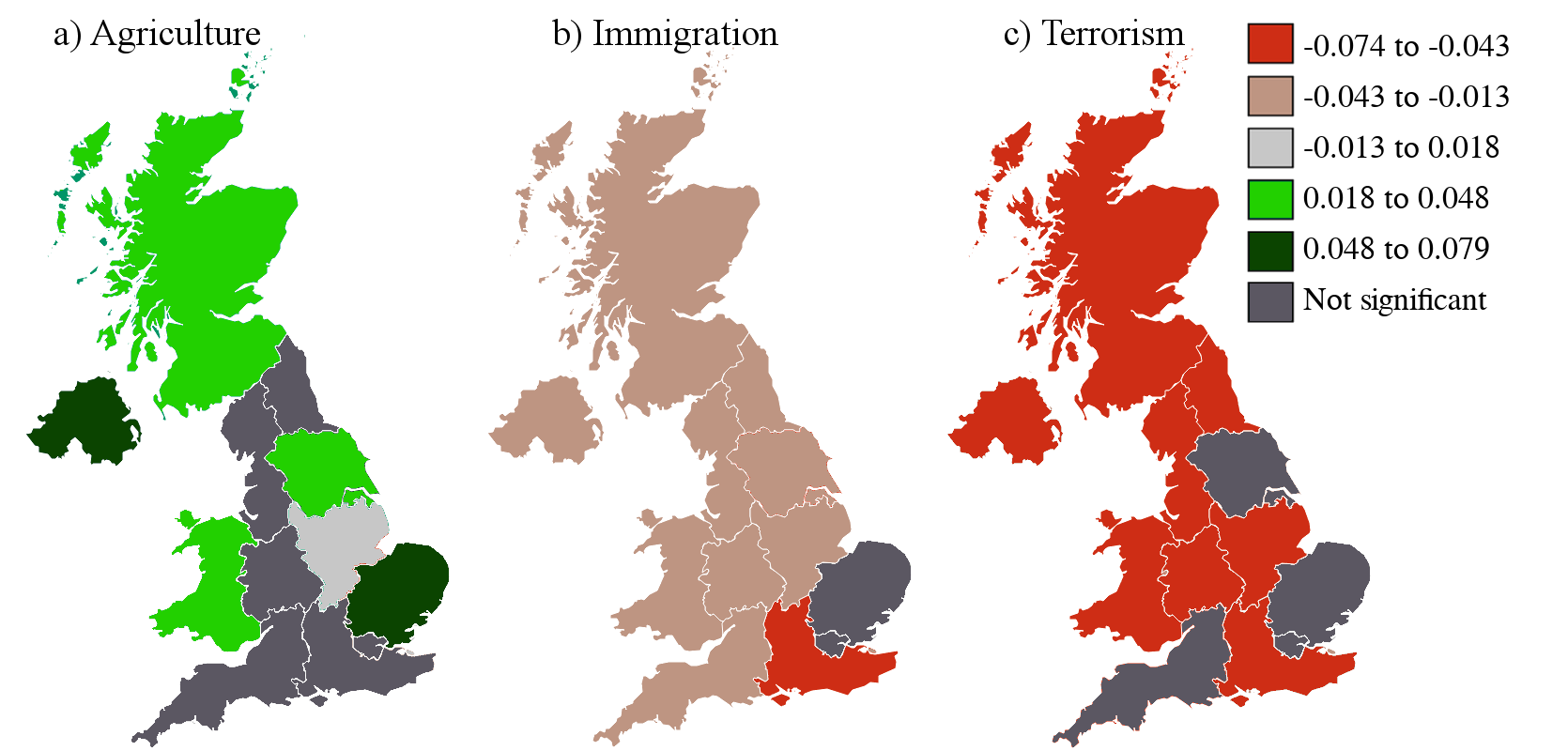}
    \caption{Entity emphasis in local vs. national newspapers. Red areas are those where the topic was mentioned more in national newspapers; green, more in local papers. All findings are significant at at least p\textless0.05 in a Mann Whitney test. Created using JSON map data shared with the authors under a CC BY license, with permission from Mark Greenwood.}
\label{fig:ents-vs-nat}
\end{figure}

We now compare local and national papers through the entities lens, in the same way as we did for topics above. For entities, this was done on a per-newspaper basis to increase reliability (topics were done on a per-article basis) with the consequence that findings were not uniformly statistically significant, and generally, entity findings are more subtle. Immigration and terrorism were mentioned generally less in local coverage, as illustrated in Fig~\ref{fig:ents-vs-nat}. Scotland was mentioned generally less in local coverage except in Scotland where it was mentioned more (p\textless0.01). Local politics were mentioned generally more in local coverage. Agriculture was mentioned somewhat more in local coverage, significantly so in Yorkshire and the Humber (p\textless0.01), East Midlands, Wales, Scotland (p\textless0.05) and Northern Ireland (p\textless0.0001). Entity findings generally support our predictions and our observations for topics. Figures are too voluminous to include in full but Table~\ref{tab:ents-all-regions} gives regional/national differences for all regional papers aggregated, though note that this glosses over important regional differences as described above, and not all regions are equally represented. Significance of sample difference is calculated on a per-newspaper basis using a Mann Whitney test.

\begin{table}
\resizebox{.65\textwidth}{!}{%
\begin{tabular}{l|l|l|l|l}
\textbf{Topic} & \textbf{Nat. av.} & \textbf{Reg. av.} & 
\textbf{p-value} & \textbf{Sig.}\\
\hline
Employment & 0.106 & 0.106 & 0.782 & -\\
Agriculture & 0.0 & 0.021 & 0.020 & *\\
Immigration & 0.06 & 0.025 & 0.000 & ***\\
Terrorism & 0.074 & 0.02 & 0.000 & ***\\
Scotland & 0.029 & 0.034 & 0.003 & **\\
EU politics & 0.177 & 0.151 & 0.004 & **\\
Trade & 0.115 & 0.097 & 0.188 & -\\
Local government & 0.0 & 0.047 & 0.000 & ***\\
NI/Wales & 0.052 & 0.038 & 0.013 & *\\
\hline
\end{tabular}%
}
\caption{Mann Whitney tests comparing average entity topic area mentions per article in national media vs local media. First two columns give the average number of mentions per article. The p-value and a visual significance indicator are then given.}
\label{tab:ents-all-regions}
\end{table}

\subsection*{RQ2: Twitter data}

%%BEGIN WITH SAMPLE FINDINGS

\begin{figure}[h!]
  \centering
    \includegraphics[width=.9\textwidth]{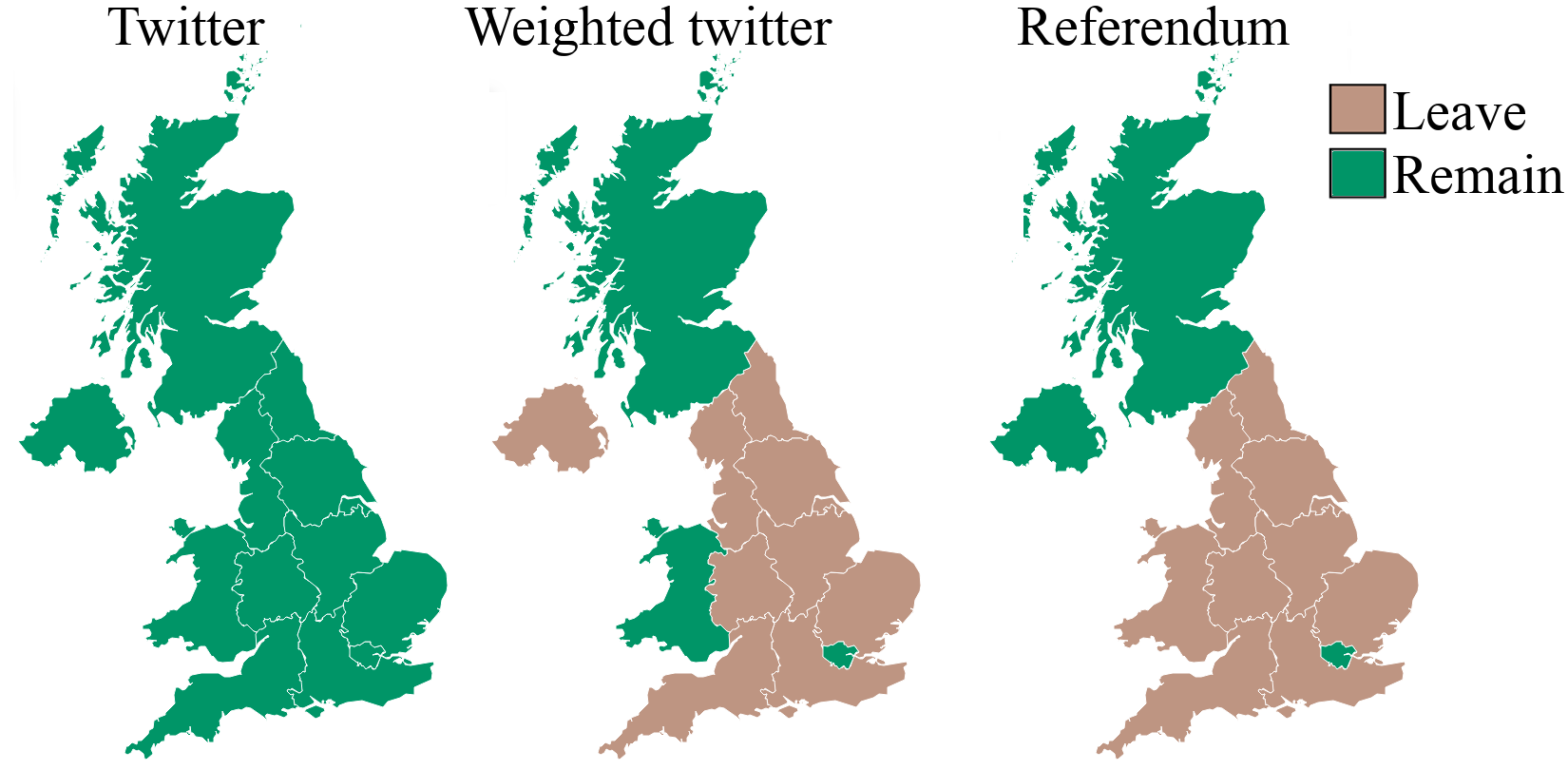}
    \caption{Leave/remain balance in Twitter corpus. Twitter users tend to be remainers. To assess how similar regional variation on Twitter is to that of the referendum, we also show the Twitter data weighted to 52\% leave. Created using JSON map data shared with the authors under a CC BY license, with permission from Mark Greenwood.}
\label{fig:sample}
\end{figure}

%%TO DO LUKE: please check these are the references you meant as I guessed them from author and date.
Research profiling Twitter users can be broadly categorized into two strands. The first gleans demographic data from users on the platform itself~\cite{barbera2015understanding,li2013spatial} whilst the second examines survey data~\cite{mellon2017twitter,greenwood2016social,duggan2015mobile}. Findings across both approaches suggest Twitter users are likely to be younger, have higher educational qualifications, be more politically engaged, and live in urban areas. Further findings point towards a left-leaning bias on the platform, and a higher prevalence of male users. Twitter activity tends to mirror mainstream politics, spiking at times of heightened political interest, such as campaign debates or scandals~\cite{jungherr2014twitter,larsson2014triumph}. Overall, Twitter is not representative of the wider population and appears to generally reflect existing inequalities in political participation, rather than act as a mechanism to flatten them~\cite{kalogeropoulos2017shares}.

Fig~\ref{fig:sample} shows the balance of remainers vs. leavers in the Twitter sample. We see that across the country, Twitter users tend to be remainers. However, we are interested to know whether the Twitter skew to remain differs by region in a way that resembles the referendum result. If we weight the Twitter sample to bring leavers up to the 52\% that voted to leave the EU in the referendum, we can get a heuristic guide. We see that on a per-region basis Twitter resembles the actual referendum outcome (RMSE 3\%, correlation of 0.89). Perhaps more interestingly however, Twitter over-represents remainers in Wales, and under-represents them in Northern Ireland. We bear this in mind as we go forward with the Twitter analysis.

Our finding that Twitter is biased toward remain fits with the picture created by previous research about the Twitter user population~\cite{barbera2015understanding}, as well as results from the survey data in Table~\ref{tab:twit-dem}, which indicate a political left bias in keeping with the remain position. Looking specifically at users who share political information in comparison to general users, education and retirement become statistically insignificant, and the left-right result weakens. However, other predictors increase in strength, suggesting gender, attention to politics, and time spent online are key predictors over and above those which predict being on the platform in the first instance. Interestingly, those categorized as C2 class (skilled working class individuals such as mechanics) were significantly less likely to share political information (-0.70, p\textless0.05), and have been suggested to be one of the more pro-Brexit sections of society (Skinner and Gottfried~\cite{skinner2017britain}, though Antonucci et al.~\cite{antonucci2017brexit} claim that the main leave voters were the ``squeezed middle'' (those with intermediate levels of education such as A levels). In the table, p\textless0.01 is signified by ``***'', p\textless0.05 by ``**'' and p\textless0.1 by ``*''. The survey data also reveals that Twitter users are more likely to be found in London than all other regions, except for East of England and Scotland which are not significantly different to the capital. This spatial variation is not found for the comparison between general users and those who share politics.

\begin{table}
\resizebox{.85\textwidth}{!}{%
\begin{tabular}{l|l|l}
 & \textbf{Twitter users (1)} & \textbf{Political sharers}\\
 & \textbf{vs Non-users (0)} & \textbf{on Twitter (1)}\\
 & \textbf{} & \textbf{vs Tw. Users (0)}\\
\hline
Age & -0.03*** & -0.01*\\
Gender (ref = Male) & -0.11* & -0.26**\\
Attended University & 0.20*** & 0.13\\
Class (ref = D/E) & &\\
- C2 & -0.16 & -0.70**\\
- C1 & 0.05 & -0.11\\
- A/B & 0.52 & -0.19\\
Retired & -0.30*** & 0.28\\
Ethnic Minority & -0.10 & 0.21\\
Daily Internet Use (ref = \textless30mins) & &\\
- Between Half hour and Hour & 0.70*** & 0.80**\\
- An hour or more & 1.13*** & 1.68***\\
Ideology (Left 0 – 10 Right) & -0.79*** & -0.18***\\
Attention to Politics: & &\\
(None 0 – 10 Great deal) & 0.07*** & 0.41***\\
\hline
\textbf{Pseudo-r2} & \textbf{0.12} & \textbf{0.17}\\
\textbf{n} & \textbf{11,995} & \textbf{3,244}\\
\hline
\end{tabular}%
}
\caption{Demographic characteristics of Twitter users: logistic regression coefficients. Approximated Social Grade classes are used: C2 covers skilled manual workers, C1 covers supervisory and professional roles and A/B covers upper management and professional roles.}
\label{tab:twit-dem}
\end{table}

%%MEDIA LINKING

\begin{figure}[h!]
  \centering
    \includegraphics[width=.9\textwidth]{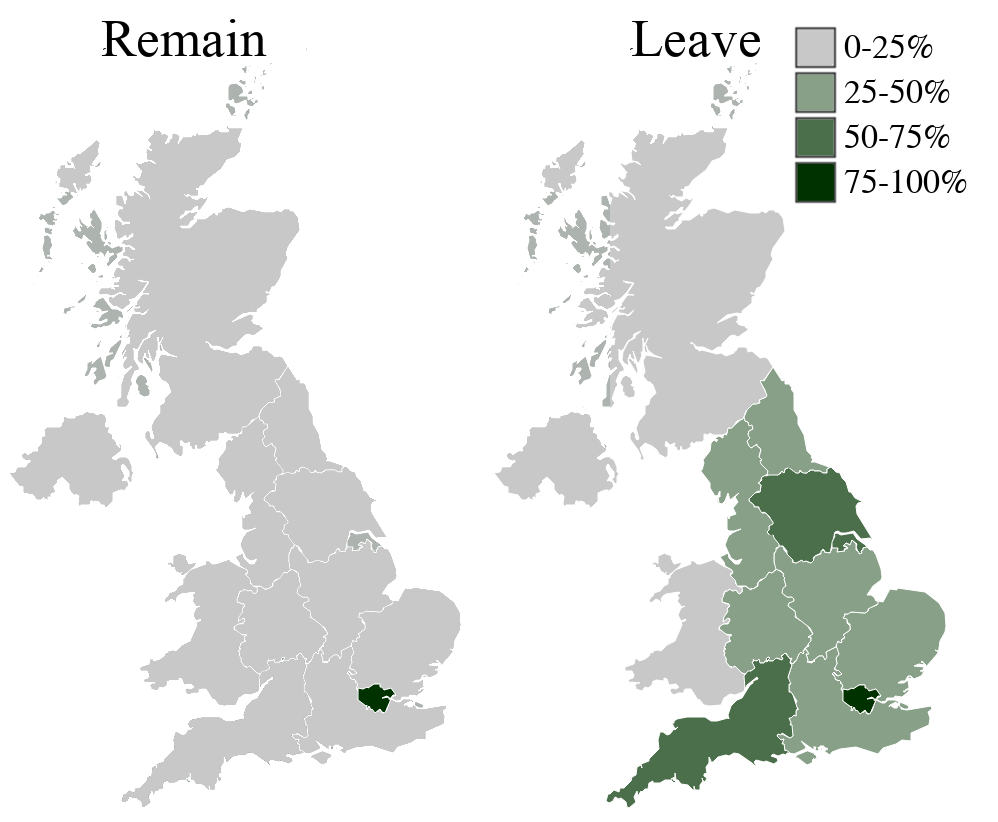}
    \caption{URLs (in tweets) per capita, split by leave/remain vote intention of the Twitter account holder. Created using JSON map data shared with the authors under a CC BY license, with permission from Mark Greenwood.}
\label{fig:links-per-capita}
\end{figure}

URLs found in the tweets were expanded from shortened forms often used on Twitter, possibly following a number of redirects, to result in a target URL. The most common newspaper/media web domains were then counted. Fig~\ref{fig:links-per-capita} shows the location from which the media link-containing tweets originated, normalized by population size of that region. In the figure, four shades indicate where the regions lie on a scale from zero to the most vocal segment (which is London leave-voters). Even after controlling for population size, London still dominates in terms of linking activity in the Brexit Twitter conversation, originating almost twice as many links per capita as any other region among leavers, more than five times as many as Northern Ireland leavers, and around four times as many links at least from remainers as any other region, a result that echoes the survey findings above with regards to location of Twitter users. Fig~\ref{fig:links-per-capita} however illustrates that the leave campaign on Twitter had markedly more engagement in other regions of England than the remain campaign, a finding that still holds for tweet count instead of link count. Further choropleths can be viewed on the project website (\url{http://services.gate.ac.uk/politics/ba-brexit/}).

Overall, the Guardian was the most linked paper across regions/voters (60,472 links, of which 67\% were from remainers), followed by the Express (56,652 links, of which 99\% were from leavers), the BBC (47,577 links of which 59\% were from leavers), the Telegraph (36,729 links, of which 83\% were from leavers), the Independent (25,645 links, of which 58\% were from remainers) and the Daily Mail (23,633 links, of which 91\% were from leavers). In most regions, the Express is the most popular link target by some margin. In the north east and south east however the Guardian edges ahead slightly. In London, Wales and Scotland, Guardian is the most popular by some margin. Across the regions, most linked sites are uniformly the Express, the Guardian, the BBC, the Telegraph, the Daily Mail and the Independent in slightly varying orders. Links to local papers are much fewer; the Yorkshire Post received 2,388 links (of which 88\% were from remainers) and the Herald, 980 links (of which 54\% were from leavers).
%These figures below are from the original Brexit corpus, so elevated by non-voters:
%Irish Times (5954 links across all regions, not in the news corpus)
%Yorkshire Post (4501 links)
%Herald (4454 links)
%Wales Online (3550 links, not in the news corpus)
%Belfast Telegraph (2658 links, not in the news corpus)
%The Scotsman (2617 links, not in the news corpus)
%Birmingham Mail (1856 links)
%Liverpool Echo (1845 links)
%Kent Online (1513 links, not in the news corpus). 

\begin{figure}[h!]
  \centering
    \includegraphics[width=.9\textwidth]{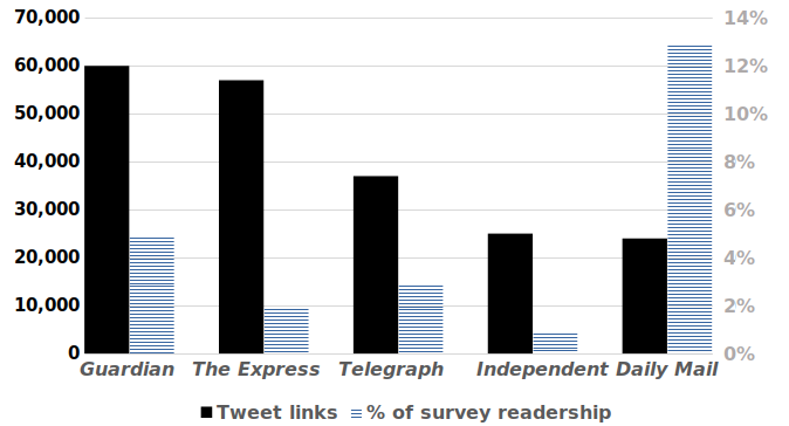}
    \caption{Tweets links from each paper analyzed in study and \% of survey readership who read paper, showing that a news source being linked on Twitter does not provide an indicator of readership.}
\label{fig:readership}
\end{figure}

Comparing these figures with the readership figures in the survey data reveals an interesting picture, as shown in Fig~\ref{fig:readership}. The media that attract the most links on Twitter are not at all the papers that have the highest circulation. The two main online influences of the Guardian and the Express are both attracting proportionally smaller numbers of people who actually state that they are readers, with the disparity in the case of the Express being particularly great. The Daily Mail, on the other hand, is reaching a much wider audience than its Twitter linking figures would suggest. This raises questions regarding the social meaning of linking to a newspaper on Twitter.

%%WHAT TWITTER IS INTERESTED IN

Subjects of general interest in the Twitter corpus (which we recall is selected according to Brexit-related hashtags so can be expected to illustrate subjects mentioned in conjunction with Brexit) include trade, immigration and football. Subjects that attracted more interest on Twitter than in the press as a whole include for example the murder of Jo Cox, the National Health Service (NHS), the rumour that Queen Elizabeth II supports Brexit, and the former UK Independence Party leader Nigel Farage. 

We saw in Fig~\ref{fig:topics-histo} that over our selection of topics, Twitter interest levels differ from those in the local and national press. Tweeters are more interested in trade, immigration and fishing than either local or national press. They are less interested in employment, local politics, steel, car pollution and terrorism. Tweeters lie in a middle ground between national (low) and local (high) levels of interest in Scotland, Northern Ireland/Wales and agriculture. All findings are significant at p\textless0.001 in a Mann Whitney test.

However, due to the high number of datapoints, the Mann Whitney test is very sensitive, and in some cases the absolute difference in mean values is small. The most notable example is between Twitter and national news interest in immigration (0.02365 vs. 0.02338) despite the apparent Twitter spike. Twitter immigration data has an odd distribution, with an unusual ``bulge'' of accounts showing medium to high levels of interest in immigration. This means that the histogram in Fig~\ref{fig:topics-histo} shows an elevated score for Twitter immigration. Other topics correspond roughly to the histogram at the start of the paper, which is calculated across the pooled material, with the exception of trade, perhaps because of the formulaic style of trade articles that may result in repeated terms. Table~\ref{tab:means} gives mean values for the different media.

\begin{table}
\resizebox{.80\textwidth}{!}{%
\begin{tabular}{l|l|l|l|l|l}
\textbf{Topic} & \textbf{Twitter} & \textbf{National} & 
\textbf{Regional} & \textbf{Leave} & \textbf{Remain}\\
\hline
Immigration & 0.024 & 0.023 & 0.02 & 0.029 & 0.02\\
NI/Wales & 0.007 & 0.005 & 0.009 & 0.006 & 0.008\\
Scotland & 0.01 & 0.01 & 0.022 & 0.01 & 0.01\\
Terrorism & 0.007 & 0.015 & 0.006 & 0.008 & 0.006\\
Steel & 0.003 & 0.007 & 0.007 & 0.003 & 0.004\\
Trade & 0.026 & 0.028 & 0.042 & 0.031 & 0.022\\
Local gov & 0.01 & 0.007 & 0.02 & 0.008 & 0.01\\
Fishing & 0.006 & 0.003 & 0.004 & 0.007 & 0.005\\
Employment & 0.013 & 0.021 & 0.034 & 0.011 & 0.015\\
Agriculture & 0.008 & 0.005 & 0.011 & 0.008 & 0.009\\
Car pollut. & 0.006 & 0.006 & 0.006 & 0.006 & 0.006\\
\hline
\end{tabular}%
}
\caption{Mean values for the topic representation in different media across articles/tweeters. Leave and remain columns average across Twitter material from accounts identified as leave or remain voters.}
\label{tab:means}
\end{table}

%PER-REGION

On a per-region basis, the only topics to show a strong correlation between local newspaper reporting and Twitter user focus are the strongly situated topics of Scotland, Northern Ireland/Wales and steel; that is to say, the extent to which these topics were discussed by tweeters in the different regions correlated significantly with the extent to which they were discussed by local papers in those regions, or by national papers in conjunction with those regions (p\textless0.001 for both leavers and remainers, except for steel/leavers where p\textless0.01). Local press showed a sensitivity to local interest levels in fishing (p\textless0.05). Interestingly, employment was reported in the national press in association with those regions where remainers \textit{weren't} talking about it (p\textless0.05), or in fact leavers particularly. The correlation is positive for local press, though not statistically significant.

In summary, Twitter data shows a number of anomalies. It arises from a biased population relative to the general UK electorate, being younger, male, educated, politically engaged and urban. It shows distortions in media popularity relative to survey readership data, for example in linking the Express far more than the Daily Mail. It shows an unusual level of interest in immigration, with an unusual distribution. However, Twitter users show less interest in terrorism than the national press, and more interest in trade, agriculture and fishing, suggesting that some Twitter use may reflect local concerns.

\subsection*{RQ3: How leavers and remainers differ}

In light of the above, we now consider the difference between leave and remain foci and the extent to which each found a voice in the press.  Immigration was more discussed by Twitter's leavers than remainers. Refugees, terrorism and fishing were more discussed by leavers. Northern Ireland was discussed more by remainers. The economy, employment, local politics and steel were more discussed by remainers on Twitter.  Scotland was equally discussed by leavers and remainers. Means are given in Table~\ref{tab:means} and all findings are significant at p\textless0.001 in a Mann Whitney test.

Whilst strong regional correlations were found in the extent of interest from newspapers (local and national) and on Twitter for the topics of the regions themselves (Scotland etc.) and for steel as described above, per-region correlations for other topics were more subtle, but in some cases present for either leavers or remainers. Coverage of local politics per-region correlated with the extent of remainers' interest as determined from the Twitter data (national press, p\textless0.05, regional p\textless0.01), and coverage of car pollution in the national press correlated with the extent of discussion by Twitter leavers in those regions (p\textless0.05).

Recall that Fig~\ref{fig:tops-reg} above shows the pattern of mentioning trade and Brexit in different regions. Trade was found to be correlated with voting leave per-region (again, p\textless0.05). The relationship between talking about trade and voting leave may reflect the initial observation that motivated the work; areas most dependent on EU trade voted to leave. Mentioning Scotland in local newspapers correlates with voting remain (p\textless0.05). No other topic among our selection showed a significant correlation, though the regional correlation between mentioning immigration and voting leave was close to significant.

A selection of newspapers was evaluated to assess whether linking to newspapers on a per-region basis correlates with leave stance and/or with remain stance. As a proportion of total links from a particular region, correlating between linking a particular paper and voting  on a per-region basis gives predictable results. Linking to the Express correlates with voting leave (-0.63, p\textless0.05), linking to the Guardian correlates with voting remain (0.74, p\textless0.01) and linking the Daily Mail correlates with voting leave (-0.74, p\textless0.01). Survey readership statistics were examined in a similar manner. Percentage of respondents who state that they read each of the following papers was ascertained, as shown in Fig~\ref{fig:readership}; the Telegraph, the Independent, the Daily Mail, the Express and the Guardian. This number alone has little meaning however for our purposes, as a 2\% readership for paper X could be high in one area but low in another where people do not tend to read a national paper so much. Therefore we totalled the readership percentages together across the five papers, per region, to give an indication of national paper readership in that area, and then took the percentage for each paper as a proportion of that. This produced interesting results. While, as we saw above, linking to the Express is a strong leave indicator, for readership the correlation is insignificant (-0.03). For the Guardian, however, the correlation is much the same as for linking (0.71, p\textless0.01). For the Daily Mail, the correlation between vote and survey readership becomes stronger (-0.80, p\textless0.002). This might suggest a geographical disconnect between the people who are linking the Express and the people who are reading it, to an extent not found in the other newspapers looked at. Recall that the Express readership is actually low, so a high Express readership compared with other regions is not incompatible with an overall remain inclination in that region. Daily Mail readership is a better indicator of a region's Brexit feelings.

Per-region correlations of local and national media topic coverage against Twitter topic interest were performed, and the resulting baskets of correlations (local vs national) were compared in a t-test. This showed that, across the eleven topics, congruence with local press is significantly greater for remainers (p\textless0.05). This provides evidence in support of the hypothesis that local awareness (or perhaps simply resistance to the national narrative) is connected to a more positive attitude toward EU membership.

\section*{Discussion}

We have presented findings from a corpus of Brexit-related tweets classified for user location and Brexit vote intent, alongside a large UK news corpus of local and national Brexit-related articles from the referendum period and a sample of survey data. We used topic modelling to ascertain differences in focus between national and local press, and how each relates to topics emerging from the Twitter data. We looked for evidence of regional sensitivity in local and national reporting, or lack thereof, and the relationship between topic focus, regional sensitivity and Brexit stance.

We find that national press emphasizes terrorism much more than local press, and immigration significantly more. Local press emphasizes trade, employment, local politics, agriculture, car pollution, fishing, Scotland and Northern Ireland/Wales. We suggest that local press appear to place greater emphasis on a range of practically relevant issues. National press show a moderate awareness of the issues affecting regions but not a high degree of sensitivity.

Twitter users show a particular interest in trade, which is associated with those wishing to leave the EU. Twitter material appears to show a high level of focus on immigration, but on inspection this is shown to be unrepresentative of Twitter users in general. The unusual distribution of users' interest levels in immigration, with a large ``bulge'' showing medium to high levels, might be suggestive of campaign activity (organized lobbying by a vocal group). Further investigation of this is important future work.

Newspapers popularly linked from tweets show a very different profile from newspapers that survey respondents say that they read. The Daily Mail was most popular among survey respondents, but on Twitter the Guardian and the Express were most popular, demonstrating that their appeal on that medium is amplified, perhaps in part due to the demographics of Twitter, but maybe also due to a difference in how information is consumed on Twitter, creating a market for stories that support different objectives such as identity signalling or influencing. The Express in particular is anomalous in that although those linking to it are almost unanimously leave voters, the number of survey respondents claiming to read the paper per region does not correlate with the overall voting pattern in that region (regions with more Express readers do not have more leave voters). This might suggest that an Express article serves a particular purpose online that it does not serve offline, and this might also be connected to the unusual Twitter activity around immigration--the Express produced an abundance of anti-immigration articles~\cite{ramsay2016monopolising}. The extent of a region's Daily Mail readership is a much better indicator of its likely Brexit stance, and indeed the lack of correlation between Express readership and regional Brexit vote might be explained by the possibility that generally speaking, the country's leavers find the Daily Mail more reflective of their attitudes.

We also find some evidence that remain voters are more aligned with the local press in terms of topic profile than leave voters. This could be seen as supporting Seaton's~\cite{seaton2016brexit} suggestion that local press are doing an important job in increasing resistance to ``propaganda and media-promoted ideas'', and in keeping with Faris et al's~\cite{faris2017partisanship} proposition of network propaganda in that to some extent Twitter sharers and some national press co-operated on an immigration and terrorism-focused narrative that didn't cover to the same extent as local press a range of practical issues relevant to people's lives.

%Seaton 2016. "Now online opinions and interests provide warm, embracing communities. Alarmingly, the message and the community have merged."

\section*{Supporting information}

\begin{table}
\textbf{\Large{Appendix}}\\
\\
\resizebox{.45\textwidth}{!}{%
\begin{tabular}{|l|l|r|}
\hline
\textbf{Newspaper} & \textbf{Region} & \textbf{Articles}\\
\hline
The Northern Echo & UKC & 405\\
Sunderland Echo & UKC & 113\\
Manchester Evening News & UKD & 169\\
Liverpool Echo & UKD & 321\\
Lancashire Evening Post & UKD & 135\\
The Bolton News & UKD & 106\\
Hull Daily Mail & UKE & 180\\
Yorkshire Post & UKE & 873\\
The Star (Sheffield) & UKE & 290\\
Yorkshire Evening Post & UKE & 192\\
York Press & UKE & 358\\
Bradford Telegraph and Argus & UKE & 67\\
Leicester Mercury & UKF & 248\\
Derby Telegraph & UKF & 313\\
Nottingham Post & UKF & 219\\
Stoke The Sentinel & UKG & 415\\
Birmingham Evening Mail & UKG & 362\\
Coventry Evening Telegraph & UKG & 225\\
Eastern Daily Press & UKH & 336\\
The Evening Standard (London) & UKI & 1571\\
Oxford Mail & UKJ & 207\\
Hampshire Chronicle & UKJ & 24\\
Bristol Post & UKK & 218\\
The Plymouth Herald & UKK & 204\\
Bournemouth Echo & UKK & 258\\
Daily Post (North Wales) & UKL & 619\\
South Wales Evening Post & UKL & 269\\
South Wales Echo & UKL & 267\\
Aberdeen Press and Journal & UKM & 509\\
Herald (Glasgow) & UKM & 1524\\
Aberdeen Evening Express & UKM & 151\\
Evening Times (Glasgow) & UKM & 154\\
Edinburgh Evening News & UKM & 190\\
Belfast Telegraph & UKN & 665\\
Irish News & UKN & 657\\
\hline
Total &  & 12814\\
\hline
\end{tabular}%
}
\caption{Regional Newspapers Article Numbers}
\label{tab:regional}
\end{table}

\begin{table}
\resizebox{.45\textwidth}{!}{%
\begin{tabular}{|l|l|r|}
\hline
\textbf{NUTS1} & \textbf{NUTS1 Name} & \textbf{Papers}\\
\hline
UKC & North East (England) & 2\\
UKD & North West (England) & 4\\
UKE & Yorkshire and The Humber & 6\\
UKF & East Midlands (England) & 3\\
UKG & West Midlands (England) & 3\\
UKH & East of England & 1\\
UKI & London & 1\\
UKJ & South East (England) & 2\\
UKK & South West (England) & 3\\
UKL & Wales & 3\\
UKM & Scotland & 5\\
UKN & Northern Ireland & 2\\
\hline
Total & & 35\\
\hline
\end{tabular}%
}
\caption{Regional Newspaper Counts}
\label{tab:regional-count}
\end{table}

\begin{table}
\begin{center}
\resizebox{.45\textwidth}{!}{%
\begin{tabular}{|l|r|}
\hline
\textbf{Newspaper} & \textbf{Articles}\\
\hline
Daily Mail and Mail on Sunday & 2041\\
Daily Star Daily Star Sunday & 328\\
Financial Times (London) & 2464\\
The Daily Telegraph (London) & 2719\\
The Guardian(London) & 9709\\
The Independent (United Kingdom) & 6833\\
The Mirror and The Sunday Mirror & 1353\\
The New Statesman & 116\\
The Observer(London) & 587\\
The Spectator & 181\\
The Sun (England) & 3050\\
The Sunday Telegraph (London) & 542\\
The Sunday Times (London) & 1336\\
The Times (London) & 4070\\
\hline
Total & 35329\\
\hline
\end{tabular}%
}
\caption{National Newspapers Article Numbers}
\label{tab:national}
\end{center}
\end{table}

\begin{table}
\resizebox{.45\textwidth}{!}{%
\begin{tabular}{|l|l|r|}
\hline
\textbf{NUTS1} & \textbf{NUTS1 Name} & \textbf{Articles}\\
\hline
UKC & North East (England) & 313\\
UKD & North West (England) & 485\\
UKE & Yorkshire and The Humber & 1384\\
UKF & East Midlands (England) & 648\\
UKG & West Midlands (England) & 548\\
UKH & East of England & 265\\
UKI & London & 2464\\
UKJ & South East (England) & 368\\
UKK & South West (England) & 584\\
UKL & Wales & 649\\
UKM & Scotland & 1104\\
UKN & Northern Ireland & 180\\
\hline
Total & & 8992\\
\hline
\end{tabular}%
}
\caption{Regional Mention Article Counts}
\label{tab:regional-mention-count}
\end{table}

\section*{Acknowledgments}

This work was supported by the UK Engineering and Physical Sciences Research Council grant EP/I004327/1 and the British Academy under call ``The Humanities and Social Sciences Tackling the UK’s International Challenges'' and by the European Union under grant agreement No. 654024 ``SoBigData''.

\nolinenumbers

% Either type in your references using
% \begin{thebibliography}{}
% Text
% \end{thebibliography}
%
% or
%
% Compile your BiBTeX database using our plos2015.bst
% style file and paste the contents of your .bbl file
% here. See http://journals.plos.org/plosone/s/latex for 

\end{document}